\documentclass[%
 reprint,
 aps,pra,
 amsmath,amssymb,
 longbibliography,
 superscriptaddress,
]{revtex4-1}

\usepackage{graphicx}
\usepackage{braket}
\usepackage{xcolor}
\usepackage{bm}
\usepackage{hyperref}

\begin{document}

\title{Crystal-momentum-resolved contributions to multiple plateaus of high-order harmonic generation from band-gap materials}

\author{Chuan Yu}
\email{chuan@pks.mpg.de}
\affiliation{Max Planck Institute for the Physics of Complex Systems, N{\"o}thnitzer Str. 38, 01187 Dresden, Germany}
\author{Hossein Iravani}
\email{hossein.iravani@modares.ac.ir}
\affiliation{Department of Physics and Astronomy, Aarhus University, 8000 Aarhus C, Denmark}
\affiliation{Department of Chemistry, Tarbiat Modares University, P.O. Box 14115-175, Tehran, Iran}
\author{Lars Bojer Madsen}
\affiliation{Department of Physics and Astronomy, Aarhus University, 8000 Aarhus C, Denmark}

\date{\today}

\begin{abstract}

We study the crystal-momentum-resolved contributions to the high-order harmonic generation (HHG) in band-gap materials, and identify the relevant initial crystal momenta for the first and higher plateaus of the HHG spectra. We do so by using a time-dependent density-functional theory model of one-dimensional linear chains. We introduce a self-consistent periodic treatment for the infinitely-extended limit of the linear chain model, which provides a convenient way to simulate and discuss the HHG from a perfect crystal beyond the single-active-electron approximation. The multi-plateau spectral feature is elucidated by a semiclassical $k$-space trajectory analysis with multiple conduction bands taken into account. In the considered laser-interaction regime, the multiple plateaus beyond the first cutoff are found to stem mainly from electrons with initial crystal momenta away from the $\Gamma$ point ($k=0$), while electrons with initial crystal momenta located around the $\Gamma$ point are responsible for the harmonics in the first plateau. We also show that similar findings can be obtained from calculations using a sufficiently large finite model, which proves to mimic the corresponding infinite periodic limit in terms of the band structures and the HHG spectra.

\end{abstract}

\maketitle

\section{\label{sec:intro}Introduction}

High-order harmonic generation (HHG) in solids has attracted great attention since its experimental observation in 2011 \cite{ghimire2011natphys}, and its underlying mechanism is still being debated \cite{ghimire2012pra,higuchi2014prl,schubert2014natphoton,luu2015nature,hohenleutner2015nature,ndabashimiye2016nature,tamaya2016prl,you2017natphys,liu2017natphys,vampa2017jpb,ghimire2019natphys}. HHG from band-gap materials can be used to produce novel vacuum ultraviolet (VUV) and extreme ultraviolet (XUV) light sources \cite{luu2015nature,garg2018natphoton}, to study the dynamics of electrons on their natural timescales \cite{hohenleutner2015nature,attosolid2018rmp}, and to investigate the properties of these materials \cite{vampa2015prl,luu2018natcommun,silva2018natphoton,silva2019natphoton,uzan2020natphoton,chacon2018arxiv}.

The solid-state HHG manifests interesting features that are quite different from the atomic and molecular cases. For example, the high-energy harmonic cutoff in a bulk crystal was found to scale quasi-linearly with the driving field strength rather than quadratically as in gas-phase atoms and molecules \cite{ghimire2011natphys}.
According to the present understanding of HHG from the class of solids, which can be adequately described by band theory (see, e.g., reviews \cite{vampa2017jpb,ghimire2019natphys}), two types of transition mechanisms are involved: ({\romannumeral 1}) interband transitions, i.e., the electron transitions between different bands at specific crystal momentum $k$, and ({\romannumeral 2}) intraband transitions which involve $k$-space electron motions within specific bands. The latter type of transitions is unique to solids and determined by the shape of band structures. According to Refs.~\cite{vampa2014prl,wang2017natcommun}, the intraband and interband HHG exhibit a fundamentally different wavelength dependence. The interplay between intraband and interband mechanisms can possibly be used to control the HHG in solids \cite{tancogne2017natcommun}.
Capturing the idea of the simple-man model for atomic HHG \cite{threestep1993prl}, a generalized three-step model \cite{vampa2014prl,vampa2017jpb} was proposed for the interband HHG in solids. It involves the following steps: ({\romannumeral 1}) Near the peak of the laser field, an electron tunnels from the valence band to the conduction band, producing a hole in the valence band; ({\romannumeral 2}) the electron and hole move in the corresponding bands, driven by the external field; and ({\romannumeral 3}) they recombine and emit HHG radiation with a frequency corresponding to the band energy difference at the crystal momentum at which the recombination occurs.

Two main theoretical approaches to study HHG in solids include ({\romannumeral 1}) the time-dependent Schr{\"o}dinger equation (TDSE) for individual Bloch electrons with specific $k$ values \cite{hawkins2015pra,wmx2015pra,wmx2016pra,bxb2016pra,bxb2017oe,dhc2017oe,liu2017pra,bxb2018pra1,jin2018jpb,jin2018pra,du2019pra}, and ({\romannumeral 2}) many-electron approaches such as the semiconductor Bloch equations (SBEs) \cite{golde2008prb,vampa2015prb,cmcdo2015pra,luu2016prb,cmcdo2017prl2,jiang2017pra,jiang2018prl,bxb2019prb,dhc2019prb,li2019prl,ly2020prl} and the time-dependent density-functional theory (TDDFT) \cite{otobe2016prb,tancogne2017prl,tancogne2017natcommun,kkh2017pra,kkh2018pra,kkh2018prl,bauer2019pra,chyu2019pra,chyu2019pra2,hoir2020prr,mrudul2020npj}.
In Ref.~\cite{dhc2018pra}, limitations of considering the response from a single electron with the TDSE in the studies of HHG in solids were discussed via comparisons with the SBEs, where the contributions from different Bloch electrons are coherently taken into account. Although focusing on the response of a single Bloch electron (e.g., at the $\Gamma$ point) has provided useful insights into the HHG mechanisms \cite{wmx2015pra,wmx2016pra}, the contributions from multiple $k$ points and their quantum interference have turned out to be important for both intraband and interband harmonics, attracting more and more attention (see, e.g., recent works \cite{bxb2019prb,li2019prl,ly2020prl}).
In addition to the methods mentioned above, the strong-field approximation (SFA) method was used to investigate HHG in solids \cite{sfa2019rpp}. A comparison study of the SFA method and the TDSE for a one-dimensional (1D) model was conducted in Ref.~\cite{jin2019pra}, where the SFA method, not accounting for the intraband current, was found to underestimate the lower-order and overestimate the higher-order harmonic yield.

In Ref.~\cite{kkh2017pra}, a finite-system 1D TDDFT model was introduced to establish the essentials of HHG by linearly polarized laser pulses in solids. With this approach, one can go beyond the single-electron models and take the electron-electron interactions into account in the TDDFT sense.
The importance of accounting for the contributions from all the orbitals in a fully-occupied band, i.e., from multiple $k$ points, was pointed out in Ref.~\cite{kkh2017pra}.
The response of a range of system sizes, from atomic to bulk solids, was studied with this model to reveal how the harmonic cutoff changes from atomic to solid-state HHG \cite{kkh2018pra}. Due to its flexibility in constructing self-consistent model systems, this methodology was applied to investigate topological edge effects \cite{kkh2018prl,bauer2019pra} and various types of imperfection effects such as doping \cite{chyu2019pra}, disorder \cite{chyu2019pra2,bauer2019pra}, and vacancies \cite{hoir2020prr}.
We note that recently such a finite-system model was also used for studying the carrier-envelope-phase effects \cite{dhc2019pra}, for examining the phase invariance of the SBEs \cite{dhc2019pra2}, and for comparing semiclassical trajectory models \cite{dhc2020pra}. For investigating HHG from an ideal periodic crystal lattice, however, it could be advantageous to extend the finite-system TDDFT model to the infinite periodic limit, as we will do in the present work.

Similar to previous studies \cite{kkh2017pra,kkh2018pra,chyu2019pra,hoir2020prr}, here we employ the 1D TDDFT model to study HHG from a linear chain of atoms exposed to linearly polarized laser pulses. In this work, we extend the previous finite-chain model to the corresponding infinitely-extended limit with a periodic treatment, and demonstrate that a sufficiently large finite model indeed mimics the infinite periodic limit in terms of band structures and HHG spectra. In terms of physical findings, we focus on the multi-plateau structure in the HHG spectra, and explain this feature with a semiclassical trajectory analysis accounting for the contributions of many independent electrons, i.e., from many different $k$ points. The infinite periodic extension of the TDDFT model offers an efficient way to simulate HHG in a perfect crystal, and makes the crystal momentum $k$ a good quantum number. We therefore investigate the contributions from electrons with different $k$ values in the HHG processes, which further support the many-electron interpretation of the multiple plateaus and provide an opportunity to identify the relevant $k$ regions for different parts of the HHG spectra. We find that electrons with initial crystal momenta located close to the $\Gamma$ point are responsible for the first plateau, while the multiple plateaus beyond the first cutoff stem from electrons with initial crystal momenta away from the $\Gamma$ point. In addition, we show that similar findings can also be obtained from the finite-system calculations.

This paper is organized as follows. In Sec.~\ref{sec:theory}, we describe the theoretical model and methods used in this work. In Sec.~\ref{sec:result}, the results of our calculations are presented and discussed. Finally, we conclude with a brief summary in Sec.~\ref{sec:concl}. The Appendix provides some details on the numerical approach used to simulate the periodic system. Atomic units (a.u.) are used throughout unless stated otherwise.

\section{\label{sec:theory}Theoretical Model and Methods}

\subsection{\label{ssec:finite_model}Finite chain model}

Let us first revisit the finite chain model and the (TD)DFT approach used in previous works \cite{kkh2017pra,kkh2018pra,chyu2019pra,hoir2020prr}.
A theoretical treatment of the corresponding infinitely extended system will be presented in Sec.~\ref{ssec:periodic_model}.
We consider a linear chain of $N$ equally charged nuclei with a separation $a$ and located at
\begin{equation}\label{eq:ionpos}
\tilde{x}_{j}=[j-(N+1)/2]a, \ (j=1,\cdots,N).
\end{equation}
The ionic potential reads 
\begin{equation}\label{eq:ionpot}
v_{\text{ion}}^{}(x) = -\sum_{j=1}^{N} \frac{Z}{\sqrt{(x-\tilde{x}_{j})^{2}+\epsilon}},
\end{equation}
where $Z$ is the charge of each ion and $\epsilon$ is a softening parameter which smoothens the Coulomb singularity in the 1D treatment.
In this work we set $Z=4$, $a=7$ and $\epsilon=2.25$, as in Refs.~\cite{kkh2017pra,kkh2018pra,chyu2019pra,hoir2020prr}.
For sufficiently large $N$, this model can qualitatively represent a bulk solid with two fully occupied valence bands.
Note that the considered model system is charge and spin neutral. Thus the number of electrons (with spin index $\uparrow, \downarrow$) is $N_{e\uparrow}=N_{e\downarrow}=2N$ for the present case of $Z=4$.
For brevity, we use the spin-restricted scheme and drop the spin index in the following formulation.

According to the Kohn-Sham (KS) theory \cite{ullrich2011book}, the field-free electronic state is described by a set of KS orbitals, determined by
\begin{equation}\label{eq:stat_kseq}
\left\{-\frac{1}{2}\frac{\partial^{2}}{\partial x^{2}} + v_{\text{KS}}^{}[n](x)\right\}\varphi_{j}^{}(x) = \varepsilon_{j}^{}\varphi_{j}^{}(x),
\end{equation}
with the static KS potential
\begin{equation}\label{eq:stat_kspot}
v_{\text{KS}}^{}[n](x) = v_{\text{ion}}^{}(x) + v_{\text{H}}^{}[n](x) + v_{\text{xc}}^{}[n](x).
\end{equation}
With $N_{\text{occ}}$ denoting the number of occupied spatial orbitals, the total density is $n(x) = 2\sum_{j=1}^{N_{\text{occ}}}|\varphi_{j}^{}(x)|^{2}$.
The Hartree potential reads
\begin{equation}\label{eq:harpot}
v_{\text{H}}^{}[n](x) = \int dx' \frac{n(x')}{\sqrt{(x-x')^{2}+\epsilon}},
\end{equation}
and the exchange-correlation potential is treated in a local density approximation (LDA)
\begin{equation}\label{eq:xcpot}
v_{\text{xc}}^{}[n](x) \simeq v_{\text{x}}^{}[n](x) = -\left(\frac{3}{\pi}n(x)\right)^{1/3}.
\end{equation}
Following previous works \cite{kkh2017pra,kkh2018pra,chyu2019pra,chyu2019pra2,hoir2020prr}, here we use the LDA exchange for the three-dimensional (3D) electron gas as a simple choice to construct our model, which aims to capture the essentials of HHG from a 3D bulk system driven by linearly polarized laser fields rather than to solve an exact 1D system.

The interaction of our model system with laser fields is simulated using TDDFT in the velocity gauge and dipole approximation. For the driving laser pulse linearly polarized along the $x$ axis, we consider the vector potential
\begin{equation}
A(t) = A_{0}\sin^{2}\left(\frac{\omega_{0}t}{2 n_{\text{cyc}}}\right)\sin(\omega_{0}t), \ (0 \leq t \leq \frac{2\pi n_{\text{cyc}}}{\omega_{0}}), \label{eq:vecpot}
\end{equation}
with $\omega_{0}$ the angular frequency (photon energy) and $n_{\text{cyc}}$ the number of cycles.
The laser-driven system is governed by the time-dependent KS equations
\begin{eqnarray}\label{eq:dyn_kseq}
&& i\frac{\partial}{\partial t}\varphi_{j}^{}(x,t) \nonumber\\
&& = \left\{\frac{1}{2}\Big({-i}\frac{\partial}{\partial x}+{A}(t)\Big)^{2} + \tilde{v}_{\text{KS}}^{}[n](x,t)\right\}\varphi_{j}^{}(x,t),
\end{eqnarray}
where the KS potential
\begin{equation}\label{eq:dyn_kspot}
\tilde{v}_{\text{KS}}^{}[n](x,t) = v_{\text{ion}}^{}(x) + v_{\text{H}}^{}[n](x,t) + v_{\text{xc}}^{}[n](x,t),
\end{equation}
is determined by the time-dependent density $n(x,t) = 2\sum_{j=1}^{N_{\text{occ}}}|\varphi_{j}^{}(x,t)|^{2}$.

The numerical calculations are performed on an equidistant grid with spacing $\Delta x=0.1$, and the number of grid points is set to $28000$ for the considered system size $N=200$.
We propagate the KS orbitals [Eq.~\eqref{eq:dyn_kseq}] using the Crank-Nicolson method with a predictor-corrector step for updating the KS potential \cite{ullrich2011book, bauer2017book}.
A fixed time step size $\Delta t \approx 0.011$ (i.e., $1/25000$ optical cycle for $2\mu$m wavelength to be considered below) is used in this work.
The convergence is further checked against time propagation using the Arnoldi-Lanczos method \cite{lanczos1986jcp} with adaptive Krylov subspaces.
The initial conditions for the TDDFT calculations, i.e., the field-free ground-state KS orbitals are found via imaginary time propagation with orthogonalization in each time step \cite{bauer2017book, itdft2019jctc}.
We note that as in previous studies \cite{kkh2017pra,kkh2018pra,chyu2019pra,hoir2020prr}, the considered laser interaction does not cause significant changes to the density; therefore we also perform time propagation with the KS potential frozen to its ground-state form, and find that the frozen KS approach captures basically the same HHG features as the dynamic KS approach. For simplicity, all the HHG spectra presented in this paper (including those for the infinite periodic system) refer to the frozen KS simulations, implying an independent-electron picture for which the contributions from different orbitals can be uniquely identified for further analysis (see below).

To calculate the HHG spectra of this model, we compute the total time-dependent current
\begin{equation}\label{eq:cur_finsys}
J_{\text{tot}}^{}(t) = \sum_{j=1}^{N_{\text{occ}}}J_{j}^{}(t),
\end{equation}
with $J_{j}(t)$ the contribution from the $j$th spatial orbital:
\begin{equation}\label{eq:cur_orb}
J_{j}^{}(t) = 2\!\int\!dx\ \text{Re}\Big[\varphi_{j}^{*}(x,t)\Big({-i}\frac{\partial}{\partial x}+A(t)\Big)\varphi_{j}^{}(x,t)\Big],
\end{equation}
where the factor of 2 accounts for the spin degeneracy.
The HHG spectral intensity is then evaluated as the modulus square of the Fourier-transformed current, i.e., 
$S(\omega)\propto \left|\int{dt}\ W(t)J_{\text{tot}}(t)e^{-i{\omega}t}\right|^{2}$, 
where $W(t)$ is a window function introduced to improve the signal-to-noise ratio.
A variety of window functions can be chosen for harmonic analysis \cite{window_harris1978ieee}. For the HHG spectra shown in this paper, we use Blackman windows with the full temporal width set equal to the laser pulse duration.

\subsection{\label{ssec:periodic_model}Infinitely extended chain model}

The above-described finite model, using a self-consistent many-electron approach, offers a straightforward way to simulate a bulk system by simply increasing the system size. In spite of its straightforwardness, the finite-size model simulations become computationally demanding for large $N$.
If one aims to simulate an ideal crystal, the numerical difficulty of the finite-size model calculations can be avoided by considering an infinitely extended system with periodicity, which is commonly used to model a bulk solid. Moreover, a periodic treatment makes the crystal momentum a good quantum number \cite{ashcroft1976book}, which could bring convenience for exploring the underlying physics.
The infinite chain limit ($N\rightarrow\infty$ with the nuclear charge $Z$ and the internuclear separation $a$ fixed) can be achieved by real-space calculations in a single unit cell with the periodic boundary condition imposed.

In the periodic treatment, we introduce the crystal momentum $k$ and the band index $m$ for the KS orbitals.
According to Bloch's theorem, we express the field-free orbitals as
\begin{equation}\label{eq:stat_bloch}
\varphi_{m,k}^{}(x) = e^{ikx}u_{m,k}^{}(x),
\end{equation}
with $u_{m,k}^{}(x)$ the periodic part $u_{m,k}^{}(x+a)=u_{m,k}^{}(x)$. 
The specific model considered here has two valence bands fully occupied ($N_{b}=2$), therefore the total density in this case is $n(x) = (2\pi/a)^{-1} \int_{-\pi/a}^{\pi/a}dk \big[2\sum_{m=1}^{N_{b}=2}|u_{m,k}^{}(x)|^{2}\big]$.

Inserting Eq.~\eqref{eq:stat_bloch} into Eq.~\eqref{eq:stat_kseq}, we obtain the equation for $u_{m,k}^{}(x)$ as
\begin{equation}\label{eq:stat_kseq_bloch}
\left\{\frac{1}{2}({-i}\frac{\partial}{\partial x}+k)^{2} + v_{\text{KS}}^{}[n](x)\right\}u_{m,k}^{}(x) = \varepsilon_{m,k}^{}u_{m,k}^{}(x),
\end{equation}
where $v_{\text{KS}}^{}[n](x) = v_{\text{ion}}^{}(x) + v_{\text{H}}^{}[n](x) + v_{\text{xc}}^{}[n](x)$ is the static periodic effective potential, and $\varepsilon_{m,k}^{}$ gives the band structure of the $m$th band.
Note that the ionic potential and the Hartree potential correspond to the infinitely extended limit of Eq.~\eqref{eq:ionpot} and Eq.~\eqref{eq:harpot}, which are unbounded with opposite signs as the system size increases, $N\rightarrow\infty$. Therefore in practical calculations, we only evaluate their sum $v_{\text{ion+H}}^{}[n](x)=v_{\text{ion}}^{}(x)+v_{\text{H}}^{}[n](x)$, which has finite values.
In our implementation, the ions are simply placed at the center of each unit cell, e.g., a unit cell $[-a/2, a/2]$ contains an ion at $x=0$.
Due to the periodicity and the charge neutrality, we can directly construct $v_{\text{ion+H}}^{}[n](x)$ from $n(x')$ in a unit cell $[-a/2, a/2]$ as \begin{equation}\label{eq:v_sum_cell_n}
v_{\text{ion}+\text{H}}^{}(x) = \int_{-a/2}^{a/2}\!\!\!\!dx' n(x') V(x,x'),
\end{equation}
with
\begin{equation}\label{eq:mat_inf_sum}
V(x,x') = \!\!\!\!\sum_{j=-\infty}^{\infty}\!\!\Big[\frac{1}{\sqrt{(x{-}x'{-}ja)^2+\epsilon}}-\frac{1}{\sqrt{(x{-}ja)^2+\epsilon}}\Big]
\end{equation}
a convergent infinite sum (for each pair of points $x$ and $x'$), which can be numerically determined before time propagation of the KS equations.

When including laser interactions in the velocity gauge, the periodicity of the Hamiltonian in space is preserved and the crystal momentum $k$ remains a good quantum number.
With the time-dependent KS orbitals expressed as $\varphi_{m,k}^{}(x,t) = e^{ikx}u_{m,k}^{}(x,t)$, the time evolution of the periodic part $u_{m,k}^{}(x,t)$ follows
\begin{eqnarray}\label{eq:dyn_kseq_bloch}
&& i\frac{\partial}{\partial t}u_{m,k}^{}(x,t) \nonumber\\
&& = \left\{\frac{1}{2}\Big({-i}\frac{\partial}{\partial x}+k+{A}(t)\Big)^{2} + \tilde{v}_{\text{KS}}^{}[n](x,t)\right\}u_{m,k}^{}(x,t), \qquad
\end{eqnarray}
with the periodic KS potential $\tilde{v}_{\text{KS}}^{}[n](x,t)$ constructed from the time-dependent density $n(x,t) = (2\pi/a)^{-1} \int_{-\pi/a}^{\pi/a}dk \big[2\sum_{m=1}^{N_{b}=2}|u_{m,k}^{}(x,t)|^{2}\big]$.

Both Eq.~\eqref{eq:stat_kseq_bloch} and Eq.~\eqref{eq:dyn_kseq_bloch} can be solved in a single unit cell $[-a/2,a/2]$ with the periodic boundary condition imposed; see the Appendix for details. Compared with the finite-system approach described in Sec.~\ref{ssec:finite_model}, the periodic treatment for the infinitely-extended model allows us to restrict the numerical calculations to a single unit cell, which significantly reduces the computational cost.
In the periodic case, we calculate the time-dependent current over the first Brillouin zone (BZ) as
\begin{equation}\label{eq:cur_cell}
J_{\text{cell}}^{}(t) = \frac{a}{2\pi}\int_{-\pi/a}^{\pi/a}\!\!\!\!dk\ J_{k}(t), 
\end{equation}
with $J_{k}^{}(t)$ the crystal-momentum-resolved current
\begin{eqnarray}\label{eq:cur_k0}
J_{k}^{}(t) = 2\!\!\sum_{m=1}^{N_{b}=2}\!\!\int_{-a/2}^{a/2}\!\!\!\!dx\ &&\textrm{Re}\Big[u_{m,k}^{*}(x,t) \nonumber\\
&& \!\!\!\!\!\!\!\!\!\!\!\!\!\!\!\!\times\Big({-i}\frac{\partial}{\partial x}+k+{A}(t)\Big)u_{m,k}^{}(x,t)\Big],
\end{eqnarray}
and again the factor of 2 accounts for the spin degeneracy.

For the real-space discretization, we use the same grid spacing as in the finite-size model calculations, i.e., a single unit cell for the considered model ($a=7$) is discretized with 70 grid points. 
Note that the Hamiltonians in Eq.~\eqref{eq:stat_kseq_bloch} and Eq.~\eqref{eq:dyn_kseq_bloch} explicitly depend on the crystal momentum $k$. Thus we use the Lanczos time propagator (with adaptive Krylov subspaces) for solving Eq.~\eqref{eq:stat_kseq_bloch} and Eq.~\eqref{eq:dyn_kseq_bloch},  which guarantees that all the KS orbitals are propagated with the same level of high accuracy. This is an advantage over some other commonly-used propagators such as Crank-Nicolson and split-operator schemes, for which the numerical error would be $k$ dependent. We use 400 equally spaced $k$ points to sample the BZ $[-\pi/a, \pi/a]$, and check the convergence against calculations using 800 BZ sampling points. All the results discussed below are fully converged.

\section{\label{sec:result}Results and Discussion}

\subsection{\label{ssec:res1}Finite versus periodic: Band structures and HHG}

The finite chain model considered here has been used in previous works for discussing many-electron effects \cite{kkh2017pra}, finite-size effects \cite{kkh2018pra}, and the influence of lattice defects \cite{chyu2019pra,hoir2020prr}. It has been demonstrated to behave like a bulk solid when the system size is sufficiently large. Since it has not been explicitly shown to what extent the finite-size system captures the features of the corresponding infinitely-extended system, here we begin our discussion with a comparison between the finite-system model and its infinite periodic limit.

\begin{figure}
\includegraphics[width=0.49\textwidth]{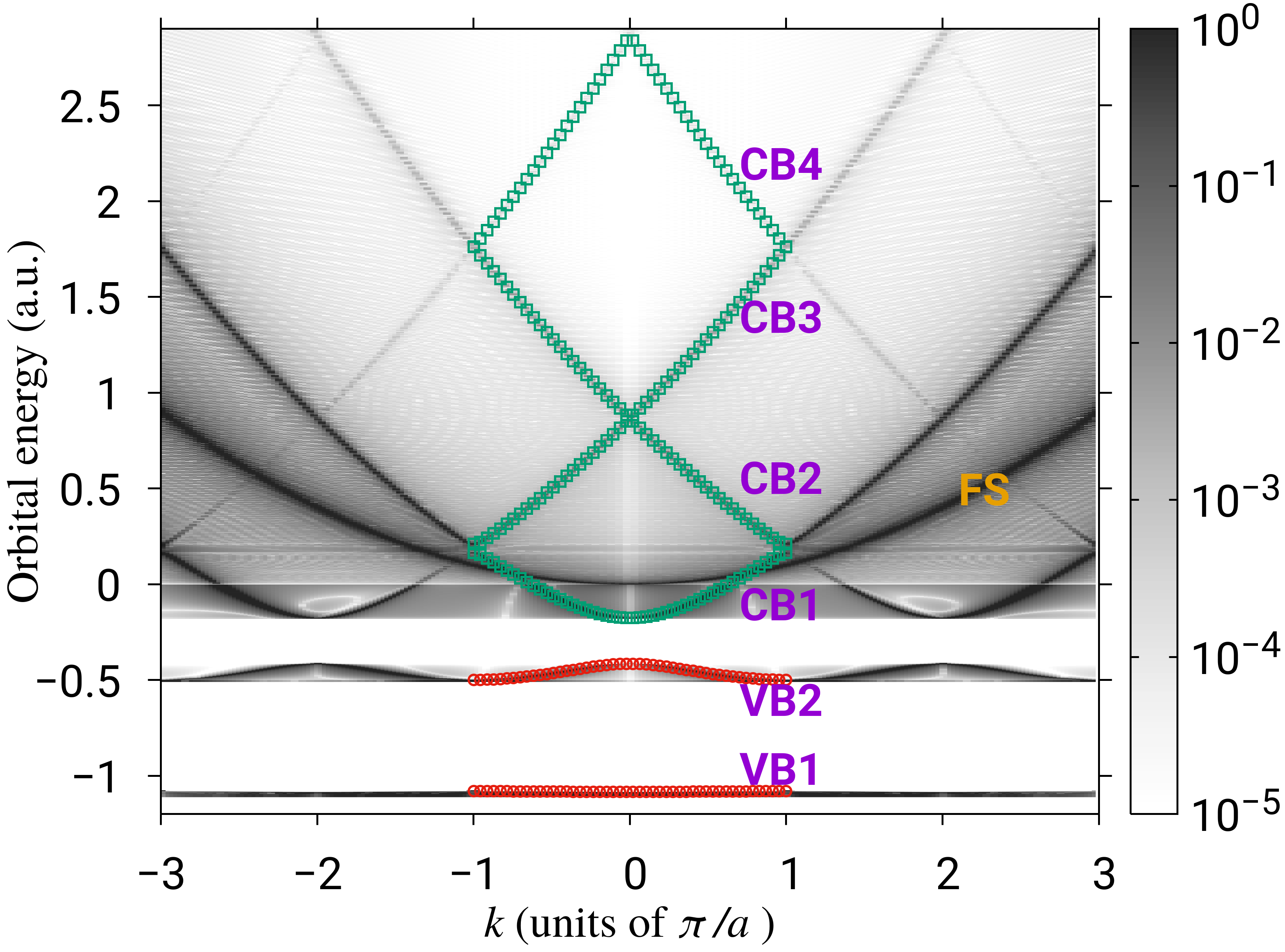}
\caption{\label{fig:fig1} Band structures for the considered finite-system model with $N=200$, obtained from the $k$-space distribution of the KS orbitals. The markers are band structures for the corresponding infinite periodic system in the first Brillouin zone. The plot includes two valence bands (VB1 and VB2, red circles) and four conduction bands (from CB1 to CB4, green squares). While the free-space (FS) dispersion $k^2/2$ naturally appears for the finite system, this finite-box feature is absent for the infinite periodic system.
}
\end{figure}

First, we present the band structures of the model in Fig.~\ref{fig:fig1}. In the figure, we include the energies for two valence bands (VB1 and VB2) and four conduction bands (from CB1 to CB4). Conceptually, the band structures are well defined for a periodic system. For a sufficiently large finite system, the $k$-space distribution of the KS orbitals can be seen as an approximate representation of the band structures. As shown in Fig.~\ref{fig:fig1}, the band structures for the $N=200$ system and the infinite periodic limit are basically identical, except that some finite-system features, i.e., the free-space dispersion $k^2/2$ and several edge states with ``out-of-band'' energies, are absent for the periodic system.
For typical laser parameters considered so far, it has been demonstrated that the orbitals in the highest-occupied band VB2 dominantly contribute to the total HHG spectra \cite{kkh2017pra}. For the considered finite system with $N=200$, the orbitals with indexes from 203 to 400 belong to VB2, and the orbitals with indexes 201 and 202 are actually a pair of edge states with energy slightly below VB2. 
Similarly, the orbitals with indexes 401 and 402 are also a pair of edge states with energy slightly below CB1. Thus in the periodic treatment, the band gap between VB2 and CB1 is found to be 0.239, which is slightly larger than the previously-reported value (0.235) \cite{kkh2017pra,chyu2019pra,hoir2020prr} obtained from the energy difference between the lowest-unoccupied and highest-occupied orbitals for the finite system.
Nevertheless, as we will see below, the above-mentioned finite-system features play negligible roles in the HHG processes for the considered model.

\begin{figure}
\includegraphics[width=0.49\textwidth]{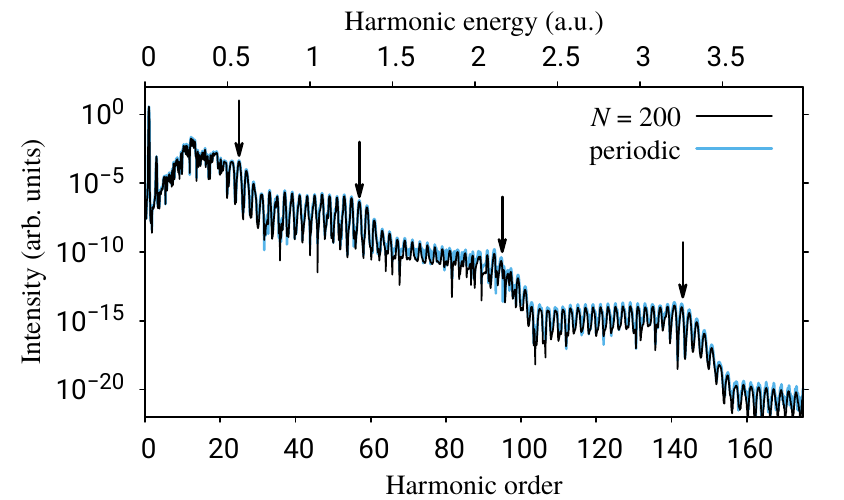}
\caption{\label{fig:fig2} HHG spectra for a finite system of $N=200$ and the corresponding infinite periodic limit. The laser parameters are $\omega_0 = 0.0228$, $A_0 = 0.24$ and $n_\text{cyc} = 15$ (see text). For the sake of comparison, the finite-system result has been scaled by $N^{-2}$ while the periodic-system result refers to the harmonics generated per atom in the infinitely-extended model. The vertical arrows indicate the cutoffs of four plateaus.}
\end{figure}

Having shown that the finite system indeed mimics the infinite periodic limit in terms of the band structures, we now compare the HHG spectra obtained from the finite system and the infinitely-extended periodic system. Since the response per atom is considered in the periodic treatment, for the sake of comparison, we evaluate the laser-driven current per atom for the finite system, and correspondingly scale the total HHG spectrum of the $N$-atom finite system by $N^{-2}$.
As one can see from Fig.~\ref{fig:fig2}, the HHG spectra obtained from the finite system and the infinite periodic system are in very good agreement, which justifies both approaches. 

For the considered laser frequency $\omega_0 = 0.0228$ (corresponding to a wavelength of ${\sim}2\mu$m), a transition from VB2 to CB1 requires at least 11 photons. This regime is similar to the one considered in experiments, where the minimum number of photons required for an excitation from the valence to the conduction band is typically between 10 and 20: A minimum number of 9 photons was required for ZnO \cite{ghimire2011natphys}, 13 for solid Kr \cite{ndabashimiye2016nature}, 15 for solid Ar \cite{ndabashimiye2016nature}, and 14--19 for bulk GaSe \cite{schubert2014natphoton, hohenleutner2015nature}.
For our model, the harmonics of orders ${\geq}11$ are therefore in the above-band-gap regime, where the interband mechanism plays an important role.
As shown in Fig.~\ref{fig:fig2}, four plateaus can be clearly identified in the spectra (with their cutoffs marked by vertical arrows), corresponding to the recombination transitions from the four conduction bands (CB$i$, $i=1,2,3,4$) to VB2. A detailed analysis of the underlying dynamics will be presented below.

\subsection{\label{ssec:res2}Multiple plateaus from a many-electron reciprocal-space perspective}

The multi-plateau spectral structure, often observed in various numerical simulations, is a signature of the presence of multiple conduction bands in a bulk solid. So far the harmonics beyond the first cutoff have been experimentally measured in solid Ar and Kr \cite{ndabashimiye2016nature}. There have been several theoretical explanations for the multiple plateaus and it seems that a consensus has not yet been reached. For example, the experimentally observed multiple plateaus were interpreted from a multilevel perspective by considering only the $\Gamma$-point ($k=0$) contribution \cite{wmx2016pra} in a model with a direct band gap at the $\Gamma$ point. Also restricting to the dynamics of a single Bloch electron at the $\Gamma$ point, a quasiclassical model was proposed, assuming that the population of higher conduction bands is pumped from the lower band step by step \cite{bxb2017oe}. Accounting for many independent electrons, a $k$-space semiclassical model was proposed, in which the band-climbing processes are also treated in a step-by-step manner \cite{ikemachi2017pra}. It should be noted that the $\Gamma$-point-only single-electron approximation was demonstrated to disagree with many-electron calculations \cite{kkh2017pra,dhc2018pra}. This point was further emphasized in a recent paper \cite{li2019prl}, where a preacceleration step was introduced for the electrons with initial crystal momenta away from the $\Gamma$ point.
For the multi-plateau structure observed in Fig.~\ref{fig:fig2}, we find that the many-electron $k$-space semiclassical model introduced in Ref.~\cite{ikemachi2017pra} can indeed provide useful insights. For completeness, we first recall the basic assumptions in this model.

(a) The laser-driven intraband motion of a Bloch electron is described by the semiclassical acceleration theorem, i.e., the instantaneous crystal momentum of an electron is $k_0{+}A(t)$, with $k_0$ the initial crystal momentum and $A(t)$ the vector potential related to the electric field $F(t)$ according to $A(t)=-\int^{t}dt'F(t')$.

(b) Tunneling excitation to an upper band is considered to occur at a band gap, e.g., at $k=0$ for the VB2-CB1 transition and at $k=\pm\pi/a$ for the CB1-CB2 transition. This is a simplifying approximation relying on the fact that the highest tunneling probability is typically obtained when the energy gap is smallest \cite{keldysh1964jetp}. In fact, transitions in the vicinity of the band gap are also possible, which we will come back to later.

(c) The interband harmonics might be emitted at any time $t$ as long as there are electrons already excited into conduction bands, and the photon energy is given by the electron-hole energy difference at the instantaneous crystal momentum $k(t)$. This is in contrast to the atomic case where recombination happens when the electron returns to the parent ion in real space. The former assumption about emission times to some extent reflects the spatially-delocalized nature of electrons in solids.
In the semiclassical picture, the electron and hole wavepackets are localized in $k$ space and spread in position over many lattice sites \cite{ashcroft1976book}, their real-space overlaps will therefore be imperfect more often than not.
Note that this description does not completely conflict with the generalized three-step model for HHG in solids \cite{vampa2014prl,vampa2017jpb}, where electron-hole recollision in real space is usually considered as a condition for the emission of harmonics. In fact, the real-space electron-hole recollision assumption will simply select the dominant contributions in some cases; more generally, the electron-hole recombination could happen even in the case of an imperfect overlap of the electron and hole wavepackets, which was quantified very recently \cite{ly2020prl}. Thus the description of harmonic emission in this $k$-space semiclassical model, albeit simple, can well capture the interband harmonics, as will be shown below.

\begin{figure}
\includegraphics[width=0.49\textwidth]{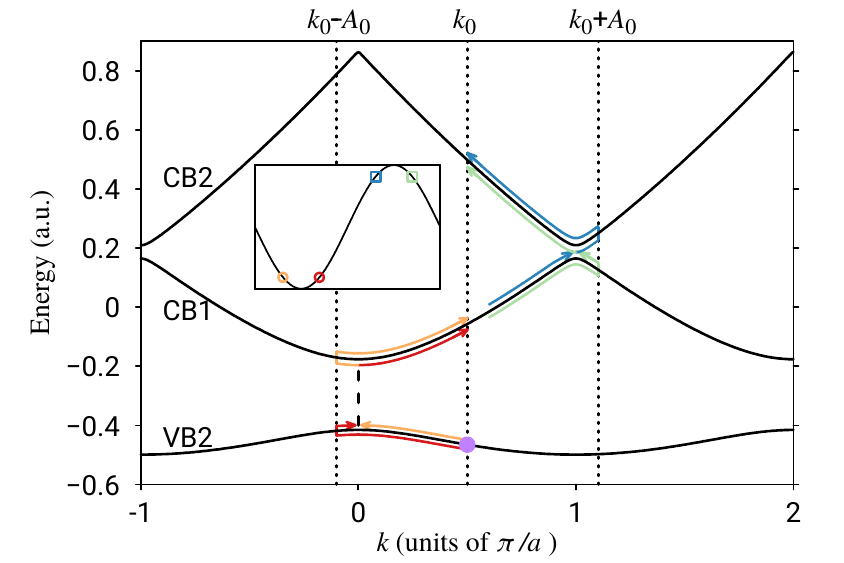}
\caption{\label{fig:fig3} The band-climbing processes illustrated by considering the central optical cycle (approximately treated as a sinusoidal form for simplicity). In the presence of the laser field, the instantaneous crystal momentum of an electron is $k_0{+}A(t)$, with $k_0$ the initial crystal momentum and $A(t)$ the vector potential (see text). The considered vector-potential amplitude $A_0$ fulfills $0.5\pi/a{<}A_0{<}\pi/a$, which applies to all the results presented in this paper. Tunneling excitation to an upper band is considered to occur at a band gap, e.g., at $k(t) = 0$ ($\pm\pi/a$) for the VB2-CB1 (CB1-CB2) transition. Due to the form of $k(t)$, there are typically two moments for a band-gap transition within one optical cycle. Similar processes can happen among higher conduction bands with additional optical cycles, e.g., after being excited into CB2, the electron may undergo a CB2-CB3 transition when $k(t)$ reaches 0 again. Inset: The vector potential in one optical cycle, with the circles (squares) indicating the moments of the VB2-CB1 (CB1-CB2) transitions for the depicted paths.
}
\end{figure}

According to this $k$-space semiclassical model, electrons can climb up to higher bands in a step-by-step manner, as illustrated in Fig.~\ref{fig:fig3}. For example, an electron with $k_0$ may undergo a VB2-CB1 transition when $k_0+A(t_1)=0$, and at a later time it continues to climb onto CB2 when $k_0+A(t_2)=\pi/a$. With additional optical cycles, climbing onto higher bands is possible, as long as $k_0+A(t)=0$ or $\pm\pi/a$ can be fulfilled.
Note that the vector-potential amplitude $A_0$ considered in this work fulfills $0.5\pi/a{<}A_0{<}\pi/a$, for which a part of electrons will reach higher conduction bands via the band-climbing processes and the HHG spectra will then show clear multi-plateau structures.
By scanning the tunneling time (denoted by $t_{s}$) for the VB2-CB1 transition and accounting for all the possible paths in $k$ space, we obtain a full set of trajectories. With the assumption of band-gap transitions, a specific electron with initial $k_0=-A(t_s)$ is automatically selected when scanning the VB2-CB1 tunneling time $t_{s}$, which implies a many-independent-electron picture. We note in passing that in this semiclassical model, the laser-induced energy shifts of bands are neglected for simplicity, i.e., the considered laser-interaction regime leads to only a small portion of electrons excited into the conduction bands, causing no significant changes to the band structures during the laser pulse.

\begin{figure}
\includegraphics[width=0.49\textwidth]{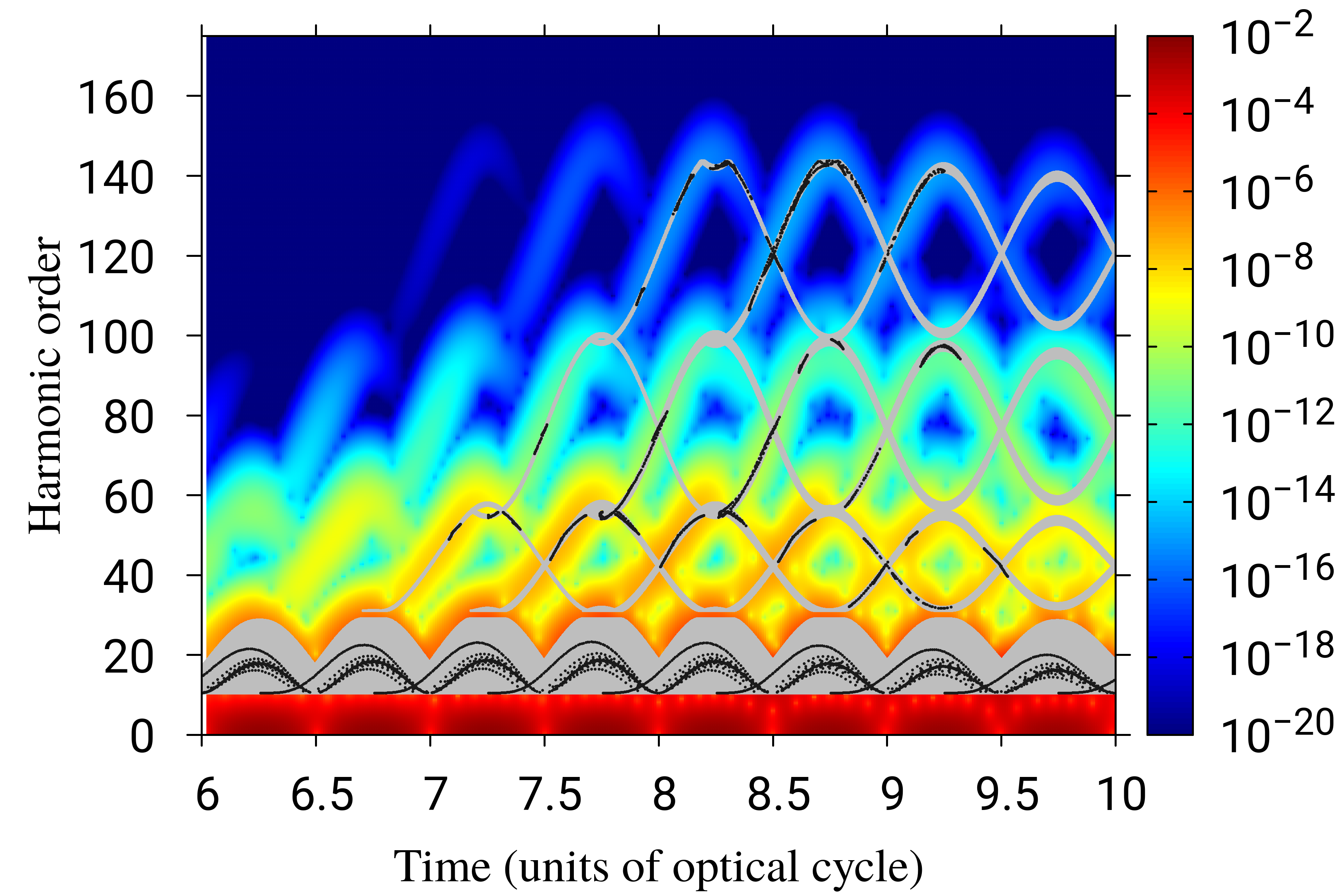}
\caption{\label{fig:fig4} Time-frequency profile of the periodic-system HHG spectrum in Fig.~\ref{fig:fig2}, extracted by a Gabor transform. The gray dots on top are obtained from $k$-space semiclassical trajectory analysis (see text). The black dots are a subset of the trajectories with the ``real-space recollision'' constraint taken into account.
}
\end{figure}

To examine the correctness of this $k$-space trajectory analysis, we compare the temporally resolved harmonic emission of the semiclassical trajectories with the time-frequency profile of the HHG spectrum. The latter, for the finite system and the infinite periodic system, is extracted by performing a Gabor transform of the time-dependent current Eq.~\eqref{eq:cur_finsys} and Eq.~\eqref{eq:cur_cell}, respectively:
\begin{subequations}\label{eq:gabor_trans}
\begin{equation}
G_{\text{tot}}^{}(\omega,t) = \int{dt'}J_{\text{tot}}^{}(t')e_{}^{-i{\omega}t'-(t-t')^{2}/(2\tau^{2})},
\end{equation}
\begin{equation}
G_{\text{cell}}^{}(\omega,t) = \int{dt'}J_{\text{cell}}^{}(t')e_{}^{-i{\omega}t'-(t-t')^{2}/(2\tau^{2})},
\end{equation}
\end{subequations}
where the width of the time window $\tau$ is chosen to be $4\pi$ (a.u.).
Figure~\ref{fig:fig4} shows the Gabor-transformed spectrum $|G_{\text{cell}}(\omega,t)|^2$ for the periodic-system current. The time-frequency profile of the finite-system HHG is very similar (not shown).
In the semiclassical trajectory analysis, the band structures of VB2 and four conduction bands (CB$i$, $i=1,2,3,4$) are taken into account, which are sufficient to describe the four plateaus observed in the spectra.
The step-by-step band-climbing processes are indeed reflected in the time-frequency profile of the HHG, i.e., the higher plateaus emerge later than the lower one.

The temporally resolved harmonic emission (i.e., the harmonic energy as a function of time) described by each $k$-space trajectory follows the instantaneous energy difference between the conduction and valence band structures, $\mathcal{E}_{cv}[k(t)]$. Note that in the semiclassical trajectory analysis with multiple conduction bands taken into account, the conduction band involved in $\mathcal{E}_{cv}[k(t)]$ is the one on which the electron moves at time $t$.
As shown in Fig.~\ref{fig:fig4}, the $k$-space trajectories, without any constraint on ``real-space recollision'' for the electron-hole pairs, are found to agree well with the time-frequency profile.
The electron-hole ``real-space recollision'' time (denoted by $t_{r}$) is typically determined by \cite{vampa2014prl,vampa2015prb,vampa2017jpb}
\begin{equation}
\int_{t_{s}^{}}^{t_{r}^{}}\!\!dt\ \nabla_{k}\mathcal{E}_{cv}[k(t)] = 0. \label{eq:eh_rec_con}
\end{equation}
{This equation, sometimes interpreted as “real-space recollision” of the electron-hole pairs, is part of the three-step model for solids and is related to the inclusion of all interband contributions along the BZ. Note that that the three-step model is formulated on the basis of spatially-delocalized Bloch waves and uses the saddle-point method to identify the significant contributions and to interpret the underlying physics \cite{vampa2014prl,vampa2015prb,vampa2017jpb}. In the semiclassical description, as noted above, the spatially-delocalized nature of the electron and hole wavepackets will inevitably cause electron-hole recombination in the cases of imperfect wavepacket overlaps, i.e., there exist a number of trajectories which do not exactly fulfill Eq.~\eqref{eq:eh_rec_con} but still contribute to the HHG. As shown in Fig.~\ref{fig:fig4}, applying the constraint of Eq.~\eqref{eq:eh_rec_con} results in a subset of trajectories covering much fewer harmonic emission events, especially in the higher-plateau region where the electron motions along higher conduction bands are involved. Therefore in the semiclassical analysis, an exact fulfillment of Eq.~\eqref{eq:eh_rec_con} seems to be a too restrictive condition, as also pointed out in Ref.~\cite{crosse2014prb}. It could be worth keeping in mind that the spatially-delocalized character of the electron-hole recombination is different from the conventional real-space picture in the atomic and molecular cases.}
Since the $k$-space trajectory analysis (without the additional ``real-space recollision'' constraint) works well for illustrating the multi-plateau HHG, here we do not go into details of the trajectories constrained by Eq.~\eqref{eq:eh_rec_con}. A detailed analysis of the ``real-space recollision'' trajectories was presented in a recent paper \cite{dhc2020pra}, with three bands (VB2, CB1 and CB2) considered. Nevertheless, it should be emphasized that the so-called ``real-space recollision'' model is simply a subset of the $k$-space trajectories. 
{Note that for a fair comparison of the $k$-space and the so-called real-space semiclassical models, one should take into account the many-electron contributions (i.e., multiple $k$ points) and the step-by-step band-climbing processes in both models. The $k$-space trajectory analysis presented in Fig.~3 of Ref.~\cite{dhc2020pra} seems not to account for these two effects, therefore a statement in Ref.~\cite{dhc2020pra} that the $k$-space trajectories disagree with the time-frequency profile could be misleading.}
Using the same TDDFT model, we show in Fig.~\ref{fig:fig4} that a properly-implemented $k$-space trajectory analysis captures the time-frequency profile of the multiple plateaus.

We also see from Fig.~\ref{fig:fig4} that the $k$-space trajectories do not cover some relatively weaker harmonic emission before the center of the pulse, e.g., do not cover the harmonics beyond the first cutoff emitted during the 7th optical cycle. This disagreement is due to the aforementioned assumption of the band-gap transitions: According to this simple approximation, the initial crystal momentum of an electron tunneling into CB1 during the early part of the pulse lies in a limited range around $k_0=0$, then the corresponding instantaneous crystal momentum $k(t)$ cannot reach the BZ edge ($\pm\pi/a$) to tunnel into CB2 for the considered $A(t)$. Accounting for tunneling transitions in the vicinity of the band gap in the semiclassical model could therefore lead to a better description.
Also, we can infer that the electrons with different crystal momenta behave differently in the band-climbing processes and may contribute to different parts in the HHG spectra. This will be further discussed in Sec.~\ref{ssec:res3}.

\subsection{\label{ssec:res3}Contribution of different orbitals to HHG}

With the insights gained from the $k$-space semiclassical model, we study how the electrons with different initial crystal momenta contribute to the HHG spectra. Recently crystal-momentum-resolved contributions were investigated for the first-plateau regime of a Kronig-Penney model in Ref.~\cite{thumm2019pra}. In the present work, we perform this type of analysis for the multi-plateau spectra of our model. Correspondingly the laser vector-potential amplitude considered here fulfills $0.5\pi/a{<}A_0{<}\pi/a$, for which it is possible to find a range of crystal momenta responsible for the high-order harmonics beyond the first cutoff.

\begin{figure}
\includegraphics[width=0.49\textwidth]{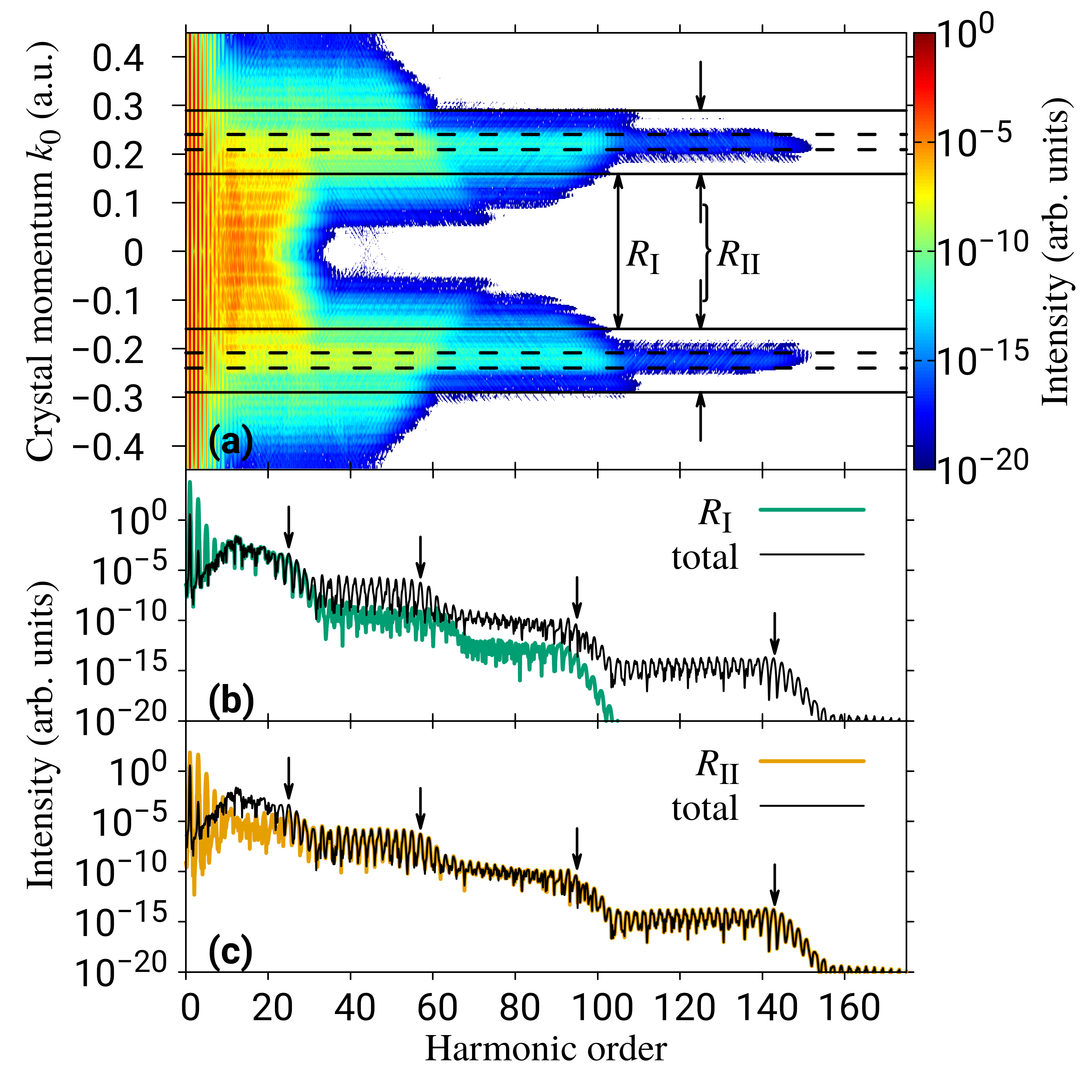}
\caption{\label{fig:fig5} (a) Crystal-momentum-resolved profile, corresponding to the periodic-system HHG spectrum shown in Fig.~\ref{fig:fig2} (see the caption of Fig.~\ref{fig:fig2} for the laser parameters). The dashed lines indicate $k_0$ values $\pm A_0$ and $\pm(\pi/a{-}A_0)$, while the solid lines indicate $k_0$ values $\pm(A_0{+}\delta_k)$ and $\pm(\pi/a{-}A_0{-}\delta_k)$. The solid lines define two regions that contribute to the majority of interband harmonics: $R_\textrm{\MakeUppercase{\romannumeral 1}} = \{k_0|\ |k_0|{\leq}(\pi/a{-}A_0{-}\delta_k)\}$ and $R_\textrm{\MakeUppercase{\romannumeral 2}} = \{k_0|\ (\pi/a{-}A_0{-}\delta_k){\leq}|k_0|{\leq}(A_0{+}\delta_k)\}$, with $\delta_k \approx 0.016\pi$ accounting for tunneling transitions in the vicinity of the band gap (see text). $R_\textrm{\MakeUppercase{\romannumeral 2}}$ captures all the harmonics beyond the first cutoff while $R_\textrm{\MakeUppercase{\romannumeral 1}}$ roughly captures the first plateau. (b) and (c) HHG spectra obtained from the current calculated within the above-defined $k_0$ regions $R_\textrm{\MakeUppercase{\romannumeral 1}}$ and $R_\textrm{\MakeUppercase{\romannumeral 2}}$, respectively. The total spectrum is also shown for comparison, with the vertical arrows indicating the cutoffs of the four plateaus as in Fig.~\ref{fig:fig2}.
}
\end{figure}

The considered 15-cycle pulse is relatively long such that the carrier-envelope-phase effects play a very minor role.
{The $k$-space semiclassical model implies that the electrons tunneling into CB1 will have initial crystal momenta $k_0$ in an interval $[-A_0, A_0]$. In pursuit of a better description, here we consider an extended region $[-A_0{-}\delta_k, A_0{+}\delta_k]$, with $\pm\delta_k$ accounting for the small $k$ range for transitions near the band gap. In this work, we roughly estimate $\delta_k \approx \sqrt{m_r \omega_0}$ with $m_r$ the electron-hole reduced effective mass. This is based on the following consideration: Among the energetically-allowed multiphoton transitions in a weak or moderate laser-interaction regime, the one involving the minimum number of photons is expected to contribute most significantly. Thus the near-band-gap transition may be roughly characterized by comparing half the photon energy (as a small width in the energy domain) with $k^2/(2m_r)$ under the effective mass approximation. For the VB2-CB1 band-gap transition in our model, we find $m_r \approx 0.11$ and thus $\delta_k \approx 0.016\pi$ for $\omega_0 = 0.0228$. Note that such an estimate is not a rigorous and general recipe; however, it is found to be applicable for our considered situation, and we imagine that it could be useful for the studies of HHG in some direct band-gap materials.
Similarly, for transitions to higher conduction bands, we have an additional condition $|k_0|{\geq}(\pi/a{-}A_0{-}\delta_k)$. (As just an approximate treatment, here we assume the same $\delta_k$ as above for simplicity, though the small $k$ region for the CB1-CB2 transition near the BZ edge could be different.)
Having the step-by-step band-climbing picture in mind, we expect the region $R_\textrm{\MakeUppercase{\romannumeral 2}} = \{k_0|\ (\pi/a{-}A_0{-}\delta_k){\leq}|k_0|{\leq}(A_0{+}\delta_k)\}$ to capture the second and higher plateaus, and the region $R_\textrm{\MakeUppercase{\romannumeral 1}} = \{k_0|\ |k_0|{\leq}(\pi/a{-}A_0{-}\delta_k)\}$ to dominantly contribute to the first plateau.}
As shown in Fig.~\ref{fig:fig5}, such an analysis is supported by the crystal-momentum-resolved profile of the periodic-system HHG spectrum, which is obtained from the Fourier transform of Eq.~\eqref{eq:cur_k0}.
Note that the notation $k$ used in Sec.~\ref{ssec:periodic_model} stands for the initial crystal momentum ($k_0$) and Eq.~\eqref{eq:cur_k0} naturally defines the contribution from a specific initial crystal momentum. 
The total spectrum is a coherent sum of all the contributions, however, the spectral intensity of individual contributions [Fig.~\ref{fig:fig5}(a)] clearly shows that the second and higher plateaus mainly originate from the region $R_\textrm{\MakeUppercase{\romannumeral 2}}$.
This is further confirmed in Figs.~\ref{fig:fig5}(b) and \ref{fig:fig5}(c) by comparing the spectra obtained from the current calculated within the above-defined $k_0$ regions [i.e., replacing the integral in Eq.~\eqref{eq:cur_cell} $\int_{-\pi/a}^{\pi/a}\!\!dk \rightarrow \int_{R_\textrm{\MakeUppercase{\romannumeral 1}(\MakeUppercase{\romannumeral 2})}}\!\!dk$]. Indeed, $R_\textrm{\MakeUppercase{\romannumeral 2}}$ captures all the harmonics beyond the first cutoff and $R_\textrm{\MakeUppercase{\romannumeral 1}}$ roughly captures the first plateau.
We note that while the region $R_\textrm{\MakeUppercase{\romannumeral 2}}$ nicely reproduces the spectrum beyond the first cutoff, it also contributes to a small part of the first plateau (see, e.g., the ``tail'' near the first cutoff at order ${\sim}25$). This is because in the step-by-step band-climbing picture, before an electron reaches higher conduction bands, it must undergo some motions within the first conduction band, partially contributing to the first plateau by CB1-VB2 recombination.

\begin{figure}
\includegraphics[width=0.49\textwidth]{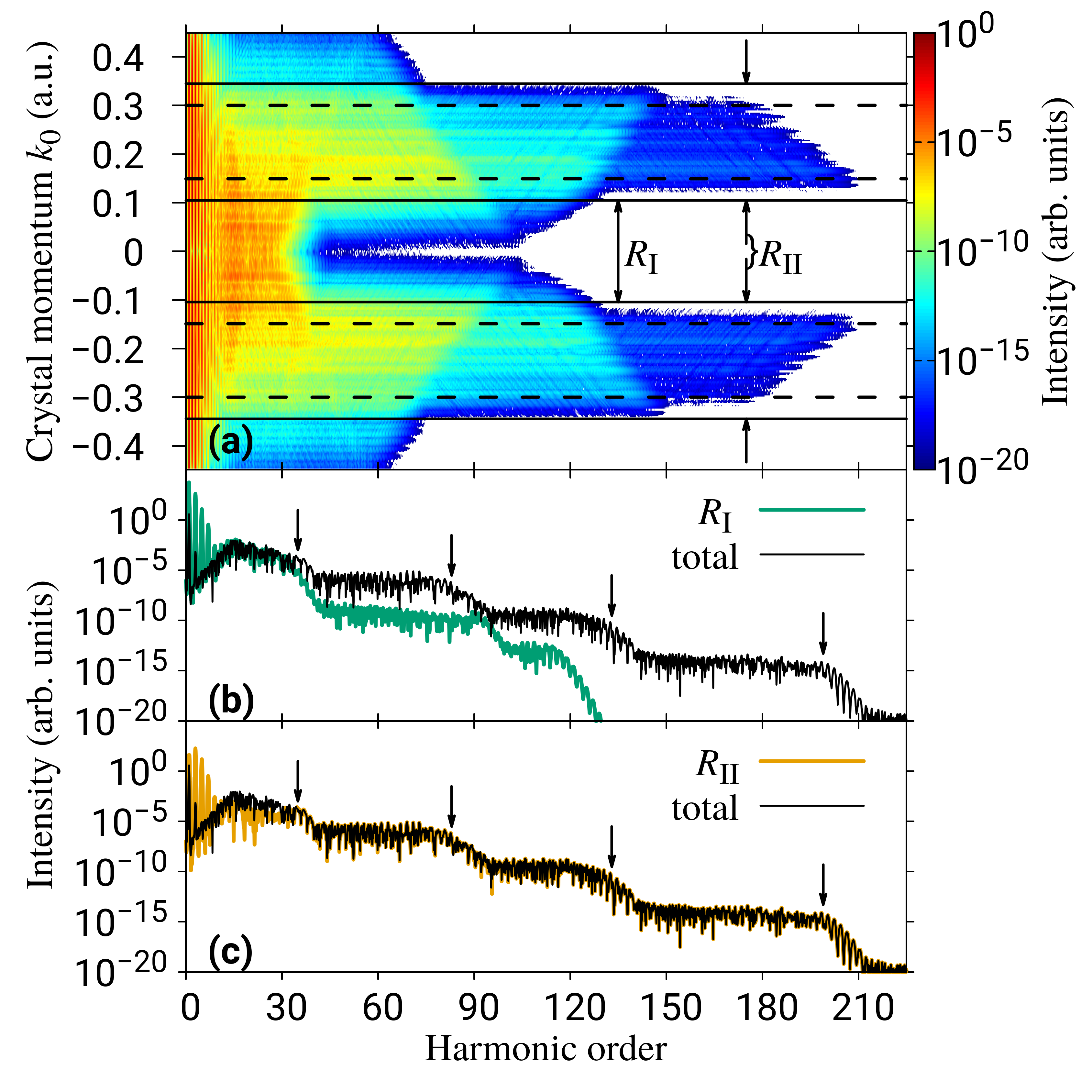}
\caption{\label{fig:fig6} Similar as Fig.~\ref{fig:fig5}, but for a different set of laser parameters $\omega_0 = 0.01824$ and $A_0 = 0.3$.
}
\end{figure}

As another demonstration of the $k_0$-resolved contributions, we present in Fig.~\ref{fig:fig6} the crystal-momentum-resolved profile for a different set of laser parameters $A_0 = 0.3$ and $\omega_0 = 0.01824$, with the peak intensity unchanged and the wavelength increased to ${\sim}2.5\mu$m. The definition of $k_0$ regions $R_\textrm{\MakeUppercase{\romannumeral 1}}$ and $R_\textrm{\MakeUppercase{\romannumeral 2}}$ also applies when varying the laser parameters. $A_0$ turns out to be an important parameter which governs the many-electron dynamics: increasing $A_0$ in the range $0.5\pi/a{<}A_0{<}\pi/a$ makes the region $R_\textrm{\MakeUppercase{\romannumeral 1}}$ ($R_\textrm{\MakeUppercase{\romannumeral 2}}$) smaller (larger). It could be interesting to note that $A_0$ is also directly linked with the Keldysh parameter \cite{keldysh1964jetp}, which typically classifies the laser-matter interaction regimes.
When increasing the wavelength at a fixed field strength, the Keldysh parameter decreases and the interband excitation becomes more deeply into the tunneling regime.
We mention in passing that increasing the wavelength while keeping $A_0 = 0.24$ fixed leads to very similar observation as Fig.~\ref{fig:fig5} except for lower HHG signals due to the weaker laser field (not shown).

\begin{figure}
\includegraphics[width=0.49\textwidth]{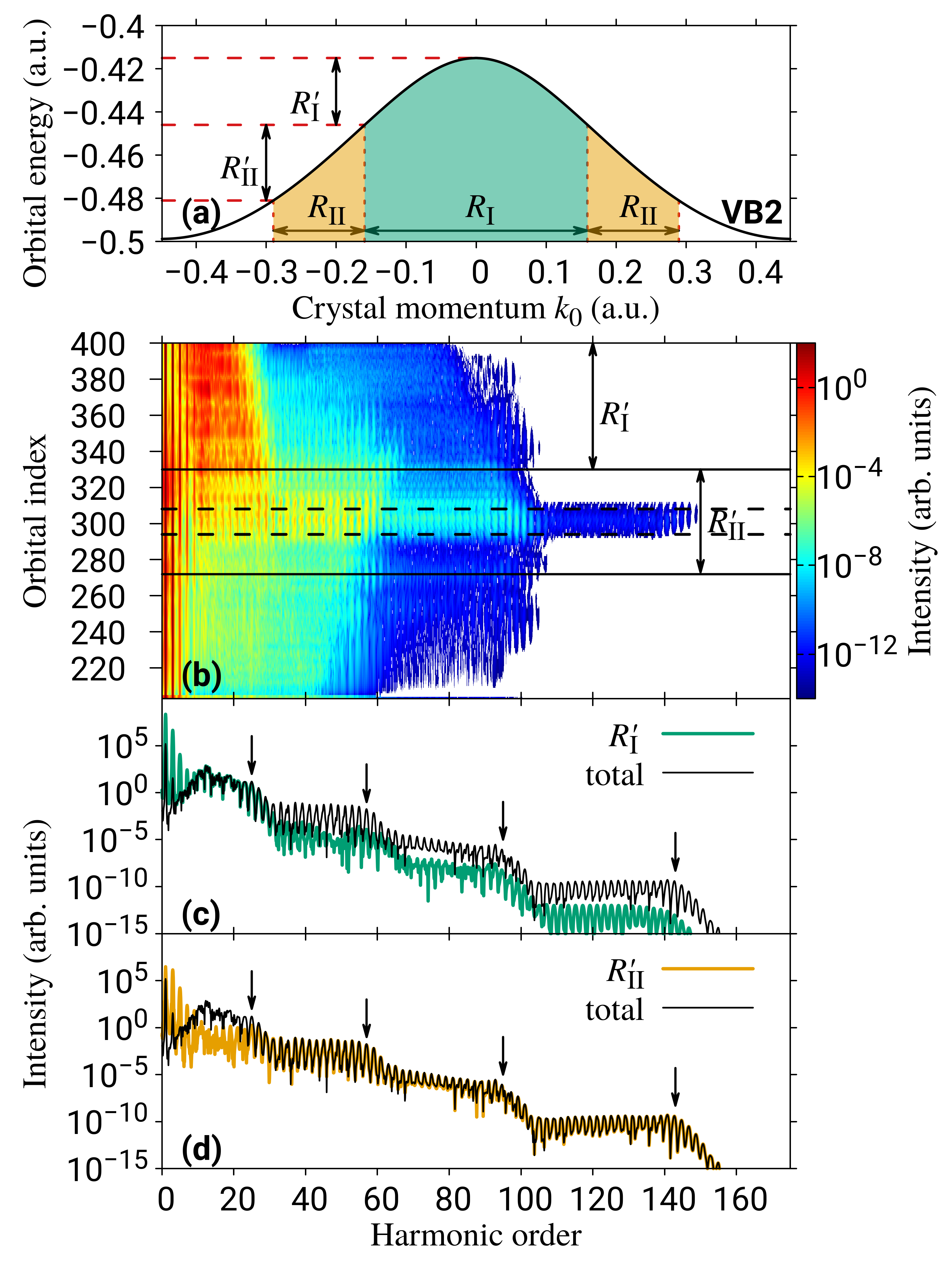}
\caption{\label{fig:fig7} (a) Mapping from the $k_0$ regions $R_\textrm{\MakeUppercase{\romannumeral 1}}$ and $R_\textrm{\MakeUppercase{\romannumeral 2}}$ defined in Fig.~\ref{fig:fig5} to the VB2 orbital energy ranges $R'_\textrm{\MakeUppercase{\romannumeral 1}}$ and $R'_\textrm{\MakeUppercase{\romannumeral 2}}$, according to the band structure. (b) Orbital-index profile, corresponding to the finite-system HHG spectrum shown in Fig.~\ref{fig:fig2} (without the scaling factor $N^{-2}$). The dashed and solid lines indicate the VB2 orbital energies corresponding to the $k_0$ values described in Fig.~\ref{fig:fig5}. For this finite system with $N=200$, the regions $R'_\textrm{\MakeUppercase{\romannumeral 1}}$ and $R'_\textrm{\MakeUppercase{\romannumeral 2}}$ are defined by indexes from 331 to 400 and indexes from 273 to 330, respectively. (c) and (d) HHG spectra obtained from the current calculated within the orbital-index ranges $R'_\textrm{\MakeUppercase{\romannumeral 1}}$ and $R'_\textrm{\MakeUppercase{\romannumeral 2}}$, respectively. The total spectrum is also shown for comparison, with the vertical arrows indicating the cutoffs of the four plateaus as in Fig.~\ref{fig:fig2}.
}
\end{figure}

In addition, we mention that as an analog to the crystal-momentum-resolved contributions for the infinite periodic model, the so-called orbital-index profile of the finite-system HHG introduced in Ref.~\cite{hoir2020prr} provides similar information when using the finite-system model. Recall that the band structures can be approximately represented by the $k$-space distribution of the KS orbitals for a sufficiently large finite system (Fig.~\ref{fig:fig1}), therefore the above-defined $k_0$ regions can be mapped, according to the valence band structure, to the orbital-index ranges for the finite-system model. The orbital-index profile of the finite-system HHG spectrum, obtained from the Fourier transform of Eq.~\eqref{eq:cur_orb}, is presented in Fig.~\ref{fig:fig7}. Here we only focus on the orbital contributions in VB2 (i.e., orbital indexes from 203 to 400), since the HHG contributions from other orbitals are almost negligible. Similar to the $k_0$ regions defined for the infinite periodic system, the orbital-index ranges $R'_\textrm{\MakeUppercase{\romannumeral 1}}$ and $R'_\textrm{\MakeUppercase{\romannumeral 2}}$ for the finite system are found to be responsible for harmonics in the first plateau and the higher plateaus, respectively.
{As discussed above, the relevant crystal-momentum (or orbital-index) ranges for interband harmonics are defined by simply considering the laser parameters and the band structures. To summarize, the relevant initial crystal momenta for capturing all the interband harmonics would be those that can be driven to the vicinity of the band gap, i.e., $[-A_0{-}\delta_k, A_0{+}\delta_k]$ in our considered case. In order to capture the harmonics in the second and higher plateau regions, we consider the condition $|k_0|{\geq}(\pi/a{-}A_0{-}\delta_k)$ to select the electrons that can reach the vicinity of the BZ edge. Note that $\delta_k$ is introduced to characterize the small $k$ region for near-band-gap transitions, which might be estimated based on the knowledge of the band structures and the laser fields. The estimate of $\delta_k$ introduced above provides a good agreement with the present numerical results. An accurate theoretical determination of $\delta_k$ would require more rigorous examination, which is beyond the scope of this work.}

Identifying relevant orbital contributions for the HHG spectral structures is helpful for understanding the underlying processes from a many-electron perspective. It could also lead to a reduction of computational cost, since one may skip the time propagation of the less important orbitals in some cases, which would be rather advantageous for 2D or 3D models.
Regarding the latter point, however, we note that while a set of significant orbitals can be defined for reproducing the interband harmonics as shown in Figs.~\ref{fig:fig5}-\ref{fig:fig7}, a correct description of the intraband harmonics in the below-band-gap regime still requires the contributions from all the occupied orbitals to be coherently taken into account, as also found in Ref.~\cite{bxb2019prb}.

\section{\label{sec:concl}Conclusion}

We numerically studied the HHG from a model solid with a band gap irradiated by linearly polarized laser pulses, in particular the multi-plateau spectral structures. We supplemented the previous model used in Refs.~\cite{kkh2017pra,kkh2018pra,chyu2019pra,hoir2020prr} with an infinite periodic treatment, and demonstrated that a sufficiently large finite model indeed mimics the infinite periodic limit in terms of the band structures and the HHG spectra.
{When discussing HHG from ideal crystals, the periodic treatment presented in this paper is more recommended than the finite-system approach used in previous works in terms of efficiency and convenience. The finite-system approach, on the other hand, has the advantage of straightforwardness and flexibility to go beyond the perfectly-periodic cases, which could therefore provide insights for manipulating HHG in the presence of topological edge states or various types of impurities and defects in solids.}

On the basis of a $k$-space semiclassical trajectory analysis, the multi-plateau spectral feature was well explained from an independent-electron perspective, where electrons with different crystal momenta contribute differently to the total HHG spectra.
It may be worthwhile to point out that the concept of semiclassical electron-hole recollisions for HHG in solids should only be considered with the spatially-delocalized nature of the Bloch waves kept in mind: An exact coincidence of the electron and hole wavepacket centers in real space could be a too restrictive condition, which cannot capture the harmonic emission in the scenario of imperfect electron-hole wavepacket overlaps.

With the infinite periodic extension of the TDDFT model, we investigated the crystal-momentum-resolved contributions in the HHG processes, and identified the relevant initial-crystal-momentum ($k_0$) regions for different parts of the HHG spectra. Specifically, for the considered laser-interaction regime, the multiple plateaus beyond the first cutoff mainly originate from electrons with $k_0$ away from the $\Gamma$ point, and the first-plateau harmonics are predominantly contributed by electrons with $k_0$ around the $\Gamma$ point.
Such particular $k_0$ ranges can be controlled by the vector potential.
Similar findings were also obtained from the finite model calculations, keeping in mind that the band structures can be approximately represented by the $k$-space distribution of orbitals for a sufficiently large finite system.
This work highlights the importance of many-electron contributions in the solid-state HHG processes. Moreover, it shows the possibility of isolating the significant contributions to some particular spectral features based on a semiclassical independent-electron picture, which lays the foundation for possible control of the HHG spectra by laser parameters to be explored in future studies.

\begin{acknowledgments}

C.Y. thanks Ulf Saalmann and Jan-Michael Rost for useful discussions.
H.I. thanks Mohsen Vafaee for helpful suggestions.
L.B.M. acknowledges support from the Villum Kann Rasmussen (VKR) Center of Excellence QUSCOPE---Quantum Scale Optical Processes and the Danish Council for Independent Research (Grant No. 7014-00092B, Grant No. 9040-00001B).

\end{acknowledgments}

\appendix*

\section{Implementation of the periodic boundary condition on a uniform grid}

In practical calculations, the wavefunctions are discretized as finite-size vectors and all the operators are expressed in terms of matrices.
In general, a detailed implementation of boundary conditions depends on the specific discretization scheme used for numerical calculations.
For completeness, here we briefly present how the 1D periodic boundary condition is implemented in this work.

The periodic treatment means that we only deal with periodic functions in practice, e.g., the periodic part of a Bloch wave which fulfills $u(x)=u(x+a)$.
To account for such a periodicity, we use a uniform $n$-point grid of size $a$ with spacing $\Delta x=a/n$.
Due to the periodicity, the grid-point positions can be chosen with an arbitrary shift; here we simply put them symmetrically in a unit cell centered at the origin $[-a/2,a/2]$: $x_{i}=[i-(n+1)/2]\Delta x,\ (i=1,\cdots,n)$.
Imagine that we apply this procedure to all the unit cells, the entire $x$ axis would then be discretized by an infinitely-extended uniform grid.
As a straightforward approach, the wavefunctions and potentials can be represented by their values on grid points; the remaining task is therefore to approximate the first- and second-order derivatives in discretized form.

Our implementation uses the matrix Numerov method \cite{matnumer2012ajp}, which is a finite-difference scheme suitable for Schr{\"o}dinger equations.
To introduce the details, let us begin with the simplest central-finite-difference method, in which the first- and second-order derivatives at the $i$th point $u'_{i}$ and $u''_{i}$ are expressed as
\begin{subequations}\label{eq:simple_fd}
\begin{alignat}{2}
u'_{i} &= \frac{u_{i+1}^{}-u_{i-1}^{}}{2\Delta x} + \mathcal{O}({\Delta x}^2), \\
u''_{i} &= \frac{u_{i+1}^{}+u_{i-1}^{}-2u_{i}^{}}{{\Delta x}^2} + \mathcal{O}({\Delta x}^2).
\end{alignat}
\end{subequations}
When restricting to the grid in a single unit cell, one should pay attention to the boundary grid points $x_{1}$ and $x_{n}$.
We denote the grid point to the left of $x_{1}$ by $x_{0}$ and the one to the right of $x_{n}$ by $x_{n+1}$, respectively.
Clearly, the periodic ansatz implies $u_{0}=u_{n}$ and $u_{n+1}=u_{1}$. 
Thus with the periodic boundary condition, one can express the first- and second-order differential operators as $n{\times}n$ matrices:
\begin{subequations}\label{eq:simple_mat_pbc}
\begin{alignat}{2}
\frac{\partial}{\partial x} &\approx \bm{\mathrm{D}}_1 = \frac{1}{2\Delta x} 
\begin{bmatrix}
 0   &   1    &   {}   &   {}   &   -1 \\
-1   &   0    &    1   &   {}   &   {} \\
{}   & \ddots & \ddots & \ddots &   {} \\
{}   &   {}   &   -1   &    0   &    1 \\
 1   &   {}   &   {}   &   -1   &    0
\end{bmatrix},\\
\frac{\partial^2}{\partial x^2} &\approx \bm{\mathrm{D}}_2 = \frac{1}{{\Delta x}^2} 
\begin{bmatrix}
-2   &    1   &   {}   &   {}   &    1 \\
 1   &   -2   &    1   &   {}   &   {} \\
{}   & \ddots & \ddots & \ddots &   {} \\
{}   &   {}   &    1   &   -2   &    1 \\
 1   &   {}   &   {}   &    1   &   -2
\end{bmatrix}.
\end{alignat}
\end{subequations}

The finite-difference method described above is for illustrative purpose only, which shows how the periodic boundary condition is imposed by setting the corner elements of the matrices.
In our practical calculations, we achieve higher accuracy [$\mathcal{O}({\Delta x}^4)$] by using an implicit three-point stencil in the spirit of the Numerov method \cite{bauer2017book}.
This elegant idea has been underlying the basics of some TDSE solvers \cite{muller1999lasphys,qprop2006cpc,solver2016cpc}.
Here we apply it to the periodic case; correspondingly, the matrix-form operators are modified as
\begin{equation}\label{eq:numerov_mat}
\frac{\partial}{\partial x} \approx \bm{\mathrm{M}}_1^{-1}\bm{\mathrm{D}}_1, \quad
\frac{\partial^2}{\partial x^2} \approx \bm{\mathrm{M}}_2^{-1}\bm{\mathrm{D}}_2,
\end{equation}
with the matrices $\bm{\mathrm{M}}_1$ and $\bm{\mathrm{M}}_2$ given by
\begin{subequations}\label{eq:numerov_mat_pbc}
\begin{alignat}{2}
\bm{\mathrm{M}}_1 &= (\bm{\mathrm{1}}+\frac{{\Delta x}^2}{6}\bm{\mathrm{D}}_2) = \frac{1}{6}
\begin{bmatrix}
 4   &    1   &   {}   &   {}   &    1 \\
 1   &    4   &    1   &   {}   &   {} \\
{}   & \ddots & \ddots & \ddots &   {} \\
{}   &   {}   &    1   &    4   &    1 \\
 1   &   {}   &   {}   &    1   &    4
\end{bmatrix},\\
\bm{\mathrm{M}}_2 &= (\bm{\mathrm{1}}+\frac{{\Delta x}^2}{12}\bm{\mathrm{D}}_2) = \frac{1}{12}
\begin{bmatrix}
10   &    1   &   {}   &   {}   &    1 \\
 1   &   10   &    1   &   {}   &   {} \\
{}   & \ddots & \ddots & \ddots &   {} \\
{}   &   {}   &    1   &   10   &    1 \\
 1   &   {}   &   {}   &    1   &   10
\end{bmatrix}.
\end{alignat}
\end{subequations}

\end{document}